\documentclass[aps,prd,showpacs,twocolumn]{revtex4}

\begin{document}

\begin{flushleft}
ADP-03-120/T558, TRI-PP-03-04, OUTP-03-13-P 
\end{flushleft}

\title{Charmonium Hybrid Production in Exclusive B Meson Decays}
 \author{F.E. Close}
\affiliation{Department of Theoretical Physics, University of Oxford, \\
Keble Rd., Oxford, OX1 3NP, England }
\author{Stephen Godfrey}
\affiliation{
Ottawa-Carleton Institute for Physics, Department of Physics  \\  
Carleton University,  Ottawa, Canada K1S 5B6 \\
{\it and} \\
Special Research Centre for the Subtatomic Structure of Matter\\
University of Adelaide, Adelaide South Australia 5000, Australia \\
{\it and} \\
TRIUMF, 4004 Wesbrook Mall, Vancouver B.C. Canada V6T 2A3}

\begin{abstract}
Recent data on charmonium production in B-meson decays suggest that
charmonium hybrid mesons with mass $\sim$4~GeV may be 
produced in $B$-decay via $c\bar{c}$ colour octet operators.  
Some of these states are likely to be narrow with clean signatures to 
$J/\psi \pi^+\pi^-$ and $J/\psi \gamma$ 
final states.  Experimental signatures and search 
strategies for existing B-factories are described.
\end{abstract}

\maketitle

\section{Introduction}

The existence of gluonic excitations in the hadron spectrum
is one of the most important unanswered questions in hadron 
physics \cite{glue}.  Hybrid mesons form one such class
which consists of a $q\bar{q}$ with an excited gluonic degree of freedom.  
Although there is mounting 
evidence for hybrids consisting of light quarks 
they still await confirmation \cite{review,beladidze,thompson}.  Recent 
observations of charmonium states in exclusive $B$-meson decays 
\cite{cleo95,belle02a,babar02,belle02b,babar02b,babar02c}
indicate that charmonium is produced significantly in the color {\bf 8}
leading us to argue that charmonium hybrids ($\psi_g$) \cite{giles77}
should also be produced in $B$-meson 
decay \cite{close98,chiladze98}.  
The unambiguous discovery of such a state would herald an 
important breakthrough in hadronic physics, and indeed, in our 
understanding of Quantum Chromodynamics, the theory of the strong 
interactions.    It would also provide important input to 
refine models of hadron structure.
In this letter we examine the production of charmonium hybrids 
and how one might observe these new types of mesons in $B$-decays.  
While various elements of our arguments have appeared elsewhere, 
by putting all the ingredients together, a more complete 
picture of charmonium hybrid production and detection emerges that, we 
hope, will encourage experimenters to pursue the necessary analysis.
In some cases we can only resort to what are at best order of magnitude
estimates of various processes.  However, we expect that they will 
provide useful guidance for the initial exploration of this new physics
frontier. 
Our goal is to point out that it may be possible to discover 
charmonium hybrids in $B$ decay and to suggest likely signatures to do 
so.
We review their spectroscopy, argue for their production in B-decays, 
and suggest experimental strategies for detecting charmonium hybrid 
mesons.

\section{Hybrid Charmonium Spectroscopy}

Lattice gauge theory and hadron models predict a rich 
spectroscopy of charmonium hybrid mesons 
\cite{giles77,isgur85,barnes95,hybrids,bernard97,perantonis90,juge97,liao02,con-glue}.  
For example,
the flux tube model predicts 8 low lying hybrid states
in the 4 to 4.2~GeV mass region with
 $J^{PC}=0^{\pm\mp}$, $1^{\pm\mp}$, $2^{\pm\mp}$, and $1^{\pm\pm}$.
Of these states the 
$0^{+-}$, $1^{-+}$, and $2^{+-}$ have exotic quantum numbers;
quantum numbers not consistent with the constituent quark model.
The flux-tube model predicts $M(\psi_g) \simeq 4 - 4.2$~GeV \cite{isgur85,barnes95};  
lattice QCD predictions for the $J^{PC}=1^{-+}$ state 
range from 4.04~GeV to 4.4~GeV \cite{bernard97,perantonis90} with
 a recent quenched
lattice QCD calculation \cite{liao02} finding 
$M(1^{-+})=4.428\pm 0.041$~GeV.   
These results have 
the $ 1^{-+}$ lying in the vicinity of the $D^{**}D$ threshold of 4.287~GeV.
There is the tantalising possibility that
the $ 1^{-+}$ could lie 
below $D^{**}D$ threshold and therefore be relatively narrow.

\section{Hybrid Production}

Recent developments in both theory and experiment lead us to
expect that charmonium hybrids will be produced in $B$ decays.  
The partial widths for $B\to c\bar{c} +X$, with $c\bar{c}$ representing 
specific final states such as $J/\psi$, $\psi'$, $\chi_{c0}$,
$\chi_{c1}$, $\chi_{c2}$, $^3D_2$, $^1D_2$ etc., 
have been calculated in the NRQCD formalism 
\cite{bodwin92,bodwin95,beneke98,ko96,yuan97,ko97} which
factorizes the decay mechanism into short (hard) and nonperturbative (soft)  
contributions.  The hard contributions are fairly well understood but 
the soft contributions, included as colour singlet and colour 
octet matrix elements, have model dependent uncertainties.
Insofar as hybrid 
$c\bar{c}$ wavefunctions have a non-trivial colour representation they 
can be produced via a colour octet intermediate state.

Over the last decade there has been great theoretical progress in the
understanding of exclusive $B$ decay to hidden charm final states
\cite{bodwin92,bodwin95,beneke98,ko96,yuan97,ko97,eichten02}.  
Central to this is the recognition of the
importance of the colour octet contributions to these decays.   
Although the colour octet terms are higher order in the velocity 
expansion for the soft contributions,
the Wilson coefficients for the colour octet subprocess are 
significantly larger than that of the colour singlet subprocess in the hard 
contributions to the decay. The net result is that the colour octet 
components play an important role in these decays.
This has been dramatically confirmed by the
observation by the CLEO, Babar, and Belle collaborations
\cite{cleo95,belle02a,babar02,belle02b,babar02b,babar02c}
of the decays  $b\to  \chi_{c0}$ and $b\to\chi_{c2}$ 
which proceed via colour octet and have been measured to have comparable 
branching fractions to the decays 
$b\to \psi +X$ and $b\to \chi_{c1}$ which have sizeable
colour singlet components.
Calculations of the BR's to the $1D(c\bar{c})$ states, which are 
higher order in the NRQCD expansion but have a
large colour octet contribution, also predict BR's 
roughly comparable to those of other charmonium states 
\cite{chiladze98,yuan97,ko97}.

The NRQCD approach assumes a Fock space expansion of the state which
gives a velocity scaling of the various wavefunction terms 
contributing to the
soft process.  For conventional mesons 
the colour singlet term is the leading term in the expansion and the 
colour octet term is higher order in $v$.  In contrast, in many models 
describing hybrid states with constituent glue, such as the bag model 
\cite{hybrids} 
and constituent gluon models \cite{con-glue}, 
the colour octet $c\bar{c}$ configuration 
is the leading term.  In the flux tube model the factorization of the 
$Q\bar{Q}$ colour is not so clear and the main uncertainty is 
estimating the hadronic matrix elements.   
Chiladze {\it et al.} \cite{chiladze98} 
estimated the octet matrix element by explicitely calculating the 
wavefunction in an approximation to the Isgur-Paton flux tube model 
\cite{isgur85} while Close \cite{close95} estimated the S-wave 
matrix element by rescaling the P-wave octet matrix element by $1/v$.
Chiladze {\it et al.} \cite{chiladze98} estimated the branching ratio 
${\cal B}[B\to \psi_g(0^{+-}) + X ] \sim 10^{-3}$ for $M \sim$ 4 GeV
(though recent quenched lattice calculations suggest $M(0^{+-})=4.70\pm 0.17$~GeV,
and hence will be inaccessible).  Note that 
the decay to the $0^{+-}$ hybrid is suppressed by a 
spin factor of $1\over 9$ while hybrid states with higher 
spin have 
larger statistical factors leading to larger branching ratios.  
Close {\it et al.}  \cite{close98} estimate a 
similar branching ratio to $1^{-+}$ and argued that if $M_g < 4.7$~GeV,
the total branching ratio
to $\psi_g$ for all $J^{PC}$ could be 
${\cal B} [\psi_g (\hbox{all }J^{PC}) +X ]\sim {\cal O}(1\%)$.  
Thus, using two different approaches for estimating 
${\cal B}[B\to \psi_g +X]$ both 
Chiladze {\it et al.} \cite{chiladze98} and 
Close {\it et al.}  \cite{close98} obtain similar results.
Both calculations estimate BR's of ${\cal O}(0.1-1\%)$ which
are comparable to the BR's for conventional $c\bar{c}$ 
states.  
We conclude that charmonium hybrids 
should be expected to be produced with roughly the same 
branching fractions as conventional charmonium states.

\section{Decays}

There are three important decay modes for charmonium hybrids:
  (i) the Zweig allowed fall-apart mode $\psi_g \to 
D^{(*,**)}\bar{D}^{(*,**)}$ \cite{ikp85,page97,page98};  
(ii) the cascade 
to conventional $c\bar{c}$ states, of the type 
$\psi_g \to (c\bar{c})(gg) \to (c\bar{c})+(\hbox{light hadrons})$
and $\psi_g \to (c\bar{c})+\gamma$
\cite{close95}; (iii) decays to light hadrons via intermediate gluons,
$\psi_g \to (ng) \to \hbox{light hadrons}$, analogous to 
$J/\psi\to  \hbox{light hadrons}$ and $\eta_c\to  \hbox{light hadrons}$.
Each mode plays a unique role.  
$\psi_g$ hybrids with exotic $J^{PC}$ quantum numbers offer the most 
unambiguous signal 
since they do not mix with conventional quarkonia.

{\it (i) Decays to $D^{(*)}D^{(*)}$:}
In addition to $J^{PC}$ selection rules (for example, 
$2^{-+}$ and $2^{--}$ decay to $D\bar{D}$ are forbidden by parity
and the exotic hybrid $\psi_g(0^{+-}) $ decays to
$D^{(*)}D^{(*)}$ final states are forbidden by $P$ and/or $C$ conservation)
a general feature 
of most models of hybrid meson decay is that decays to two mesons with 
the same spatial wave function are suppressed \cite{page97b}.  Therefore,
decays to $D\bar{D}$ should not occur, but 
small differences in wavefunctions could lead 
to small but finite widths to $DD^*$.
Seeing $DD^*$ but not $D\bar{D}$ or $D^*\bar{D}^*$ 
would be a striking signal for a hybrid meson.
The dominant coupling of charmonium hybrids is to excited states, in 
particular $D^{(*)}(L=0)+D^{**}(L=1)$ states
for which the threshold is $\sim 4.3$~GeV.  This 
is at the kinematic limit for most mass predictions so 
that decays into the preferred $D^{(*)}D^{**}$ states 
are expected to be significantly suppressed if not outright 
kinematically forbidden.
Estimates for the various $\psi_g$ decay widths and branching 
ratios are given in Table I \cite{page98}.   In columns 1-4 a 
hybrid mass of 4.1~GeV is assumed, which is below $DD^{**}$ threshold, 
while column 5 and 6 employs a mass of 4.4~GeV, above $DD^{**}$ threshold.  
This enables us to gauge the model dependence of the results and the 
effect of opening up the $DD^{**}$ channel on the total width.
The original Isgur Kokoski Paton flux tube model \cite{ikp85} 
predicts partial 
widths of $\sim 1-20$~MeV, depending on the $J^{PC}$ of the hybrid 
\cite{page97} 
while a refined version of this model predicts smaller partial widths 
of 0.3-1.5~MeV  \cite{page98}.  These widths are
quite narrow for charmonia of such high mass.
If the hybrid masses are 
above $D^{**}$ threshold then the total widths increase to 4-40~MeV for 
4.4~GeV charmonium hybrids which are still relatively narrow.  
The challenge is 
to identify decay modes that can be reconstructed by experiment.

{\it (ii) Decays to $(c\bar{c})+(\hbox{light hadrons})$:}
The 
$\psi_g \to (c\bar{c})+(\hbox{light hadrons})$ 
mode offers the cleanest signature for $\psi_g$ observation if its
branching ratio is large enough.
In addition, a small total width also offers the possibility that the 
radiative branching ratios into  $J/\psi$, $\eta_c$, $\chi_{cJ}$, and 
$h_c$
could be significant and offer a clean signal for the detection of 
these states.

For masses below $D D^{**}$ threshold the 
cascade decays
$\psi_g \to (\psi,\; \eta_c, \ldots) +(gg)$ and annihilation decays
$\psi_g (C=+) \to (gg) \to \hbox{light hadrons}$
will dominate.  If the masses of exotic $J^{PC}$ states are 
above  $D D^{**}$ threshold their widths are 
also expected to be relatively narrow for states of such high mass, 
in which case
cascades to conventional $c\bar{c}$ states transitions of the type 
$\psi_g \to (\psi,\;\psi')+(\hbox{light hadrons})$
should have significant branching ratios \cite{close95}
making them important signals to look for in $\psi_g$ searches.
The hadronic transition rates to conventional charmonia from either 
$C=+$ or $C=-$ are similar
because both are the same order in $\alpha_s$ 
and the charge conjugation of the conventional $c\bar{c}$
daughter should be the same as that of the $c\bar{c}$ hybrid parent
since two 
gluons $(C=+)$ are emitted in the lowest order process.  

A transition between two quarkonium states proceeds via the emission 
of gluons by the heavy quark and the subsequent conversion of the
gluons into light hadrons \cite{yan80}.  
The emission of the gluons is typically 
treated as a multipole expansion of the colour gauge field to estimate 
rates for hadronic transitions between $Q\bar{Q}$ states.  Kuang and 
Yan \cite{kuang81} estimated the matrix elements between quarkonium 
states by inserting intermediate states with the string in its vibrationally 
excited lowest mode, ie. hybrid states.  Thus, the matrix elements for 
hadronic  transitions between conventional quarkonia are related to 
hybrid-conventional quarkonium hadronic transitions. 
The widths $(c\bar{c}) \to \pi\pi J/\psi$ are typically 
${\cal O}(10-100)$~keV 
while $(c\bar{c}) \to \eta J/\psi$ are typically ${\cal O}(10)$~keV 
\cite{eichten02}.  
(BES has recently measured $\Gamma(\psi(3770)\to J/\psi 
\pi^+\pi^-)=(139\pm 61 \pm 41)$~keV \cite{bes} which is a $^3D_1\to 
^3S_1 \pi \pi$ transition.)
It seems reasonable to assume that the
partial widths for the decays 
$\psi_g(1^{-+}) \to \eta_c +(\pi\pi, \eta, \eta') $ and
$\psi_g(0^{+-},2^{+-})\to J/\psi +(\pi\pi,\eta, \eta')$
will be similar 
in magnitude, of ${\cal O}(10-100)$~keV.
Clearly this is not a rigorous result but it does offer a reasonable 
order of magnitude estimate.

While there are no calculations for radiative transitions involving 
charmonium hybrids there are estimates of radiative 
transitions involving hybrids with light quarks 
\cite{page95,close03}.  Both calculations found that the $E1$ transitions 
between hybrid and conventional states to be comparable in magnitude 
to transitions between conventional mesons. 
While neither calculation can be applied directly to $c\bar{c}$ 
one might take this to suggest that 
the partial widths for 
$\psi_g(1^{-+}) \to \gamma + (J/\psi, h_c)$ 
and $\psi_g (0^{+-},2^{+-}) \to \gamma + (\eta_c, \chi_{cJ})$
are the same order of magnitude 
as transitions between conventional 
charmonium states.  However, it is not at all clear if this extrapolation 
to charmonium hybrids is correct as in the flux-tube 
model Close and Dudek 
\cite{close03} showed that the $\Delta S=0$ E1 transitions to hybrids 
only occur for charged particles, and hence would vanish for 
$c\bar{c}$.
The $\Delta S=1$ M1 transitions can occur, but 
are non-leading and less well defined.
Estimates \cite{close03} for their widths are ${\cal O}(1-100)$~keV. 
Clearly, given our general lack of understanding of radiative transitions 
involving hybrids, the measurement of these transitions, 
$\psi_g \to (c\bar{c}) \gamma$, has important implications for model 
builders.

{\it (iii) Decays to light hadrons:}
Decays of the type $\psi_g  \to \hbox{light hadrons}$ offer the 
interesting possibility of producing light exotic mesons.
Estimates of annihilation widths to light hadrons will be order of 
magnitude guesses at best due to uncertainties in wavefunction effects 
and QCD corrections.  We estimate the annihilation widths 
$\Gamma[\psi_g (C=-) \to \hbox{ light hadrons}]$
and $\Gamma [c\bar{c}(C=+) \to \hbox{ light hadrons}]$ by 
comparing them to $\Gamma(\psi' \to  \hbox{ light hadrons})$ and 
$\Gamma (\eta_c' \to \hbox{ light hadrons})$.
The light hadron production rate from $\psi_g(C=-)$ decays 
is suppressed by one power of $\alpha_s$ with respect to $\psi_g(C=+)$ 
decays.  This very naive assumption gives
$\Gamma[\psi_g (C=-) \to \hbox{ light hadrons}]\sim {\cal O}(100)$~keV
and $\Gamma [c\bar{c}(C=+) \to \hbox{ light hadrons}]\sim {\cal 
O}(10)$~MeV \cite{gr02}.  These widths could be 
smaller 
because the 
$q\bar{q}$ pair in hybrids 
is expected to be separated by a distance of order 
$1/\Lambda_{QCD}$ resulting in a smaller annihilation rate than the 
$S$-wave $\psi'$ and $\eta_c'$ states.

\begin{table*}
\caption{Decays of exotic Charmonium Hybrids. 
The fall-apart widths are taken from Page, 
Swanson and Szczepaniak \cite{page98} where IKP refers to the Isgur 
Kokoski Paton model and PSS to the Page Swanson Szczepaniak model.
$D^{**}(1^+_L)$ and $D^{**}(1^+_H)$ are the low and high mass $1^+$ 
charmed meson states.  The high mass state is identified with the 
$D_1(2420)$ state. The hadronic cascade decays are rough estimates.  
See discussion in the text.
}
\begin{center}
\begin{tabular}{llrrrrrr}
\hline
$\psi_g$ State & Final State  & \multicolumn{2}{c}{IKP}  & 
		\multicolumn{2}{c}{PSS} & \multicolumn{2}{c}{PSS} \\
		&	&  \multicolumn{2}{c}{$M(\psi_g)=4.1$~GeV} 
		& \multicolumn{2}{c}{$M(\psi_g)=4.1$~GeV} 
		& \multicolumn{2}{c}{$M(\psi_g)=4.4$~GeV} \\
	&	& Width (MeV) & BR (\% ) 
		& Width (MeV) & BR (\% ) 
		& Width (MeV) & BR (\% )  \\
\hline
$1^{-+}$ & $D^*D$ & 4 & 28.4 & 0.8 & 7.3 & 0.1 & 0.3 \\
	& $D^{**}(2^+)D$ & - & - & - & - & 0.5 & 1.3 \\
	& $D^{**}(1^+_H)D$ & - & - & - & - & 25 & 63.5 \\
	& $D^{**}(1^+_L)D$ & - & - & - & - & 3.7 & 9.4 \\
	& $\eta_c \pi \pi$ & 0.1 & 0.7 & 0.1 & 0.9 & 0.1 & 0.3 \\
	& light hadrons & 10 & 70.9 & 10 & 91.7 & 10 & 39.4 \\
	& $\Gamma_{Total}$ & 14.1 &  & 10.9 &  & 39 & \\
\hline
$2^{+-}$ & $D^*D$ & 1 & 83.3 & 0.3 & 60 & 0.2 & 5.1 \\
	& $D^{**}(2^+)D$ & - & - & - & - & 0.5 & 12.8 \\
	& $D^{**}(1^+_H)D$ & - & - & - & - & 3 & 76.9 \\
	& $J/\psi \pi \pi$ & 0.1 & 8.3 & 0.1 & 20 & 0.1 & 2.6 \\
	& light hadrons & 0.1 & 8.3 & 0.1 & 20 & 0.1 & 2.6 \\
	& $\Gamma_{Total}$ & 1.2 &  & 0.5&  & 3.9 & \\
\hline
$0^{+-}$& $D^{**}(1^+_L)D$ & - & - & - & - & 25 & 62.2 \\
	& $D^{**}(1^+_H)D$ & - & - & - & - & 15 & 37.3 \\
	& $J/\psi \pi \pi$ & 0.1 & 50 & 0.1 & 50 & 0.1 & 0.2 \\
	& light hadrons & 0.1 & 50  & 0.1 & 50  & 0.1 & 0.2 \\
	& $\Gamma_{Total}$ & 0.2 &   & 0.2 &  & 40.2 &  \\
\hline
\end{tabular}
\end{center}
\end{table*}

\section{Signatures}

The decays discussed above lead to a
number of possible signals:   $\psi_g \to 
D^{(*)}D^{(*,**)}$, $\psi_g(0^{+-},2^{+-}) \to J/\psi + 
(\pi^+\pi^-,\eta,\eta')$, 
$\psi_g(1^{-+}) \to \eta_c  +  (\pi^+\pi^-,\eta,\eta')$,  
$\psi_g \to (c\bar{c}) \gamma$, 
and $\psi_g \to \hbox{light hadrons}$.
Of the possible decay modes,  $\psi_g \to J/\psi \pi^+\pi^-$,
$\psi_g \to J/\psi \eta$, and $\psi_g\to (c\bar{c})\gamma$ 
give distinctive and easily reconstructed signals.  In the former case
the subsequent decay, $J/\psi\to e^+e^-$ and $\mu^+\mu^-$ offers 
a clean tag for the event.  

A good place to search for hybrids in $\psi_g \to J/\psi \pi^+\pi^-$ 
is to look for 
peaks in the invariant mass distributions $M(e^+e^- \pi^-\pi^+)-M(e^+e^-)$.
Babar  observed a strong signal for the $\psi'$ 
in such a  distribution from the
decay chain $B\to \psi(2S) +X \to J/\psi(e^+e^-)\pi^+\pi^-$.
Babar's efficiency for the  $\psi(e^+e^-)\pi^+\pi^-$
final state is about 20\%.
With $2.3\times 10^7$ $B\bar{B}$ pairs from 
an integrated luminosity of 20.3 fb$^{-1}$ Babar observed
$\simeq 972 $ $\psi(2S)\to J/\psi \pi^+\pi^-$ events.  
Both the Babar and Belle collaborations have collected over
100 fb$^{-1}$ of integrated luminosity   
and each expects to collect 400 fb$^{-1}$ over the next few years.

Both the $0^{+-}$ and $2^{+-}$ should decay via the $\psi_g\to J/\psi 
\pi\pi$ cascade.  
Although the lattice predictions for the $0^{+-}$ and $2^{+-}$ masses are 
above $DD^{**}$ threshold there is still considerable uncertainty in 
these values and the flux tube model predicts masses approximately 
4.1~GeV.  We therefore consider both cases, where the $\psi_g$ lies 
both below and above $DD^{**}$ threshold.  For the low mass scenario, 
combining our estimates of ${\cal B}(B\to \psi_g +X) \simeq 10^{-3}$ 
and 
${\cal B}[\psi_g(2^{+-}) \to J/\psi \pi^+\pi^-] \simeq 0.2$ 
(the 4.1~GeV PSS case in Table I)
with the PDG value of
${\cal B}(\psi\to \ell^+ \ell^-) = 11.81\% $ \cite{pdg} and the Babar detection 
efficiency we estimate that for 100~fb$^{-1}$ of integrated luminosity
each experiment should 
observe roughly 50 events.  If the $2^{+-}$ lies above the $DD^{**}$ 
threshold the BR for $2^{+-}\to J/\psi \pi\pi $ decreases 
significantly to 
$2.6\times10^{-2}$ (the 4.4~GeV PSS case in Table I)
lowering the expected number to about 6 events.
Similarly,  for the $0^{+-}$ hybrid we estimate roughly 1200 events if 
it lies below threshold but only 5 events once the $DD^{**}$ decay 
modes open up.  

The $1^{-+}$ state is expected to be the lightest exotic $c\bar{c}$ 
hybrid \cite{bernard97,liao02}
and therefore the most likely to lie below $DD^{**}$ threshold. 
However, in this case the cascade goes to $\eta_c \pi\pi$, a more 
difficult final state to reconstruct.
We use our estimate of ${\cal B}(B\to \psi_g +X) \simeq 10^{-3}$ and 
the estimate given in Table I for the 4.1~GeV PSS case of
${\cal B}(\psi_g(1^{-+} \to \eta_c \pi^+\pi^-) \simeq 9\times10^{-3}$.
The Babar collaboration studied the decay $B\to\eta_c K$ by observing 
the $\eta_c$ in $KK\pi$ and $KKKK$ final states.  Combining the PDG 
values for the BR's to these final states with the Babar detection 
efficiencies of roughly 15\% and 11\% respectively 
we estimate that for 100~fb$^{-1}$ each experiment should 
observe roughly 10 events. If the $1^{-+}$ lies above the $DD^{**}$
threshold, the BR for $1^{-+}\to \pi\pi \eta_c$ decreases to 
$3\times10^{-3}$ lowering the expected number to about 3 events.

The radiative transition, $\psi_g (1^{-+})\to \gamma J/\psi$, also has a 
distinct signal if it has a significant branching ratio.
If we take the conservative value of 
$\Gamma(\psi_g(1^{-+}) \to \gamma J/\psi)\simeq 1$~keV,  
the BR's for this transition would be rather small.  On the other hand, a 
monochramatic photon offers a clean tag with a high efficiency.  One 
could  look for peaks in $M(\mu^+\mu^- 
\gamma) - M(\mu^+\mu^-)$.   Babar observed  $\chi_{c1}$ 
and $\chi_{c2}$ this way \cite{babar02} obtaining
$\simeq 394$ $\chi_{c1}$'s and $\simeq 1100$ $\chi_{c2}$'s
with a 20.3 fb$^{-1}$ data sample and an
efficiency of about 20 \% for the $J/\psi \gamma$ 
final state \cite{babar02}.  So although the rate may be too 
small to observe, given the potential payoff, it is probably worth 
the effort to perform this search.  We also note that both 
Babar and Belle should be able to see the $^3D_2$ state via 
radiative transitions \cite{eichten02} so that even if they do not discover a 
charmonium hybrid they will almost certainly add to our knowledge of 
quarkonium spectroscopy.

We note that 
Babar has measured BR's for $B\to (c\bar{c})+h$ 
at the $10^{-6}$ level with $61.6 \times 10^6$ $B\bar{B}$ pairs
$({\cal B}(B^+ \to \chi_{c0} K^+,\; \chi_{c0} \to \pi^+ \pi^- )
= (1.46 \pm 0.35 \pm 0.12) \times 10^{-6}$) \cite{babar03} demonstrating 
the accessability to these levels of combined BR's.

Experiments might also look for charmonium hybrids in invariant mass 
distributions of light hadrons.  For example,
Belle observed the $\chi_{c0}$ by looking at the invariant mass 
distributions from the decays $\chi_{c0}\to \pi^+\pi^-$
and $\chi_{c0}\to K^+K^-$ \cite{belle02a}.  They found 
efficiencies of 21\% for  $\chi_{c0}\to \pi^+\pi^-$
and 12.9\% for $\chi_{c0}\to K^+K^-$, obtaining $\sim 16$ events in 
the former case and $\sim 9$ in the  latter.  

The decay to charmed mesons also needs to be studied.  
Because there are more particles in the final state 
it will be more difficult to reconstruct the charmonium hybrid. 
On the other hand, with sufficient 
statistics these channels will be important for measuring the $\psi_g$ 
quantum numbers and distinguishing their properties from conventional 
$c\bar{c}$ states.

The final consideration in charmonium hybrid searches is distinguishing 
the signal from background.  
The largest backgrounds in these final states will be via charmonium 
states with similar masses and fairly narrow total widths 
\cite{eichten02}.  
The only charmonium states with these properties are the missing 
$c\bar{c}$ $^3D_2$ and $^1D_2$ states whose masses are predicted to be 
$\sim 3.8$~GeV \cite{gi}. 
They lie  
below $D\bar{D}^*$ and are
forbidden to decay into $D\bar{D}$ because of parity conservation.  So 
they are expected to have narrow widths of 300-400 keV and should 
be easily tagged through the dominant E1 transitions into the 
$\chi_{cJ}$ states.  These states
are expected to be produced with branching ratios of ${\cal 
O}(1\%)$ with ${\cal B}[\psi(2^{--})\to \pi\pi J/\psi]\simeq 0.12$ 
\cite{chiladze98,yuan97,ko97}.  
The successes of QCD motivated quark models 
for these states give reasonably reliable predictions for their masses 
so it should be possible to distinguish them from the charmonium 
hybrid on this basis.  In any case, their discovery would be 
interesting in their own right.
For exclusive processes such as $B\to \psi_g + K^{(*)}$ the 
$K^{(*)}$ would have a definite momentum in the $B$ rest frame.  
Careful study of $K^{(*)}$ momentum spectra is another tool that 
could be used to separate the signal from other sources, and seek excess of
low momentum $K^{(*)}$ recoiling against the $\psi_g \sim 4-4.5$~GeV.

\section{Conclusions}

Establishing the existence of mesons with explicit gluonic degrees of 
freedom is one of the most important challenges in strong 
interaction physics.  As demonstrated by the discovery of the 
$\eta_c(2S)$ in $B$ decay, $B$ decays offer a promising approach to 
discovering charmonium hybrid mesons.  
In this letter we have described a strategy to search for these states. 
While there is no question that our estimates for the various partial 
widths are crude, the essential point is that these states are expected 
to be relatively narrow and that distinctive final states are likely 
to have observable branching ratios.  Given how much we can learn by 
finding these states we strongly advocate that some effort be devoted 
to their searches.

\noindent
{\it Note added:} After the completion of this paper the Belle 
Collaboration published the observation of a new charmonium state in 
exclusive $B^\pm \to K^\pm \pi^+ \pi^- J/\psi$ decays with mass 
$3871.8\pm 0.7 \hbox{(stat)} \pm 0.4 \hbox{(syst)}$~MeV which is most 
likely the $^3D_{c2}$ state \cite{belle03}.
However, if this state were found to have natural spin parity, $0^+$,
$1^-$, $2^+$ etc,
then a dynamical suppression, such as expected for hybrids, would be called 
for.

\acknowledgments

The authors thank J. Dudek for helpful communications.
SG thanks the DESY theory group for their hospitality where some of 
this work took place.
This research was supported in part by the Natural Sciences and Engineering 
Research Council of Canada, the UK PPARC
 and European Union program ``Euridice" HPRN-CT-2002-00311.

\end{document}